\documentstyle[epsfig,times]{article}
\include {header}
\begin{document}
\begin{titlepage}

\samepage{
\setcounter{page}{1}
\rightline{CERN--TH--2000/208}
\rightline{ACT-10/00}
\rightline{CTP-TAMU-22/00}
\rightline{\tt hep-th/0007115}
\rightline{July 2000}
\vfill

\begin{center}
 {\Large \bf Results from an Algebraic Classification }\\
\vspace{.05in}
{\Large \bf of 
Calabi-Yau Manifolds }\\
\vspace{.1in}

\vfill
 \vspace{.1in}
 {\large F. Anselmo$^{1}$, J. Ellis$^{2}$, D.V. Nanopoulos$^{3}$
   $\,$and$\,$ G. Volkov$^{4}$\\}
 \vspace{.25in}
 {\it $^{1}$ INFN-Bologna, Bologna, Italy\\}
 \vspace{.05in}
 {\it  $^{2}$ Theory Division, CERN, CH-1211 Geneva, Switzerland \\}
 \vspace{.05in}
 {\it  $^{3}$ Dept. of Physics,
 Texas A \& M University, College Station, TX~77843-4242, USA,  \\
 HARC, The Mitchell Campus, Woodlands, TX~77381, USA, and \\
 Academy of Athens, 28~Panepistimiou Avenue,
 Athens 10679, Greece\\}
 \vspace{.05in}
 {\it  $^{4}$
Theory Division, Institute for High-Energy Physics, Protvino, Russia\\}

\vspace{.25in}

{\bf Abstract}


\end{center}


We present results from an inductive algebraic approach to the systematic
construction and classification of the `lowest-level' CY$_3$ spaces 
defined as zeroes of polynomial loci associated with reflexive
polyhedra, derived from suitable vectors in complex projective
spaces. These CY$_3$ spaces may be sorted into `chains' obtained by
combining lower-dimensional projective vectors classified previously. We
analyze all the 4~242 (259, 6, 1)  two- (three-, four-, five-) vector
chains, which have, respectively, K3 (elliptic, line-segment, trivial) 
{sections}, yielding 174~767 (an additional 6~189, 1~582, 199)  distinct
projective vectors that define reflexive polyhedra and
thereby CY$_3$ spaces, for a total of 182~737. These CY$_3$
spaces span 10~827 (a total of 10~882) distinct pairs of Hodge numbers
$h_{11}, h_{12}$. Among these, we list explicitly a total of 212
projective vectors defining three-generation CY$_3$
spaces with K3 {sections}, whose characteristics we provide. 



\vfill
\smallskip }

\end{titlepage}

\section{Introduction}

One of the preferred roads towards string phenomenology is via
compactification on a Calabi-Yau (CY) manifold~\cite{CY}, either of the
perturbative weakly-coupled heterotic string~\cite{quartet}, or in
some strongly-coupled
incarnation such as $M$ or $F$ theory~\cite{MF}. Most interest has
centered on CY
manifolds with three complex dimensions (CY$_3$), but their
four-dimensional CY$_4$ relatives furnish interesting compactifications
of $F$ theory, and their two-dimensional K3 relatives are also
interesting for illustrative studies. A powerful systematic approach to
the construction of CY spaces has been made possible by Batyrev's
formulation of them as toric varieties in weighted complex projective
spaces, that may be associated with reflexive polyhedra~\cite{Bat}. 
This work has opened the way for a {\it Calabi-Yau Genome
Project}~\cite{Ibanez} to
classify our possible phenomenological heritage from string theory. In
this spirit, an enumeration
of all the 473~800~776 reflexive polyhedra that exist in four dimensions,
together with a scatter plot of the 30~108 distinct pairs of the Hodge
numbers $h_{12}, h_{11}$ of the corresponding CY$_3$ spaces, has been
provided by Kreuzer and Skarke~\cite{KS}. 

An informative way to construct such reflexive polyhedra algebraically is
to obtain them from vectors in complex projective spaces~\cite{AENV}.
These may be
built up systematically as discrete linear combinations of vectors in
complex projective spaces of lower dimensions, in a way that
is universal for all dimensions and naturally gives
base-fibre structures. This approach furnishes
additional insights into duality, yields relations between spaces in
different dimensions obtained by simple geometrical intersection and
projection operations, and leads naturally to the appearance of
Cartan-Lie algebra structures associated with singularities of the
corresponding CY spaces~\cite{enhanced,Ietal}. It also provides
illuminating
relations between
CY spaces in complex projective spaces with the same number of dimensions,
that appear in chains of different
discrete linear combinations of lower-dimensional vectors. This
constructive approach also provides automatically information about
fibrations~\cite{AENV}, in particular the elliptic and K3 fibrations of
CY$_3$ spaces
that are of interest in $M$ and $F$ theory~\cite{MF}.  

We have previously illustrated the value added by this systematic
algebraic approach to the explicit construction of CY spaces in different
dimensions by a discussion of K3 spaces~\cite{AENV}.  In particular, we
constructed
95 `lowest-level' K3 spaces as zeroes of the polynomial loci associated
with projective vectors, and a further 730 as `higher-level'
intersections of such polynomial loci. We also used our algebraic
construction to display the corresponding reflexive polyhedra and to
illustrate their duality~\cite{duality}, intersection and projection
properties, and the
association
of Cartan-Lie algebra symmetries with singularities of these K3 spaces.

In this paper, we demonstrate the power of our technique in the
interesting case of CY$_3$ spaces, reporting the algebraic
constructions of all the spaces obtainable at the lowest level as simple
polynomial loci. As a first step, we have surveyed all the CY$_3$ spaces
that belong to the 4~242 chains, identified in our previous
paper~\cite{AENV}, that
can be generated by pairs of projective vectors extended from lower
dimensions. Among the 174~767 lowest-level CY$_3$ spaces found this way,
there are 94 distinct projective vectors, with related reflexive
polyhedra, that define CY$_3$ spaces with $h_{11}
- h_{12} = N_g$, the number of generations, equal to three.
We also find 118  distinct projective vectors corresponding to spaces
with $h_{11} - h_{12} = -3$, whose mirrors~\cite{mirror} are
three-generation CY$_3$
spaces. Among these 212 = 94 + 118 three-generation
CY$_3$ spaces there is just one mirror pair. By
construction, all of these CY$_3$ spaces have K3
{bases}~\cite{KLM,KV,AKMS,CPR}.
We tabulate the Hodge numbers of these three-generation
CY$_3$ spaces and the numbers of vertices and edges of the
associated reflexive polyhedra and their mirrors, of which there are
179 distinct octuples.

We then go on to explore the CY$_3$ spaces obtainable from the 259
three-vector chains also identified in our previous paper~\cite{AENV}.
These chains
yield 6~189 additional lowest-level CY$_3$ spaces, none of which have
$N_g = 3$. All of these spaces, by construction, possess elliptic
{sections}.  We complete our construction of the lowest-level CY$_3$
spaces with those obtainable from the six four-vector chains: 1~582
additional CY$_3$ spaces, and the one
quintuple-vector chain: 199 additional CY$_3$ spaces. Neither of these
sets of chains contribute any new $N_g = 3$ spaces.

In total, this lowest-level algebraic construction produces 10~882 of the
30~108
distinct pairs of CY$_3$ Hodge numbers $h_{12}, h_{11}$ found by Kreuzer
and Skarke~\cite{KS}, and enables us to understand several features of
their
`scatter plot' of Hodge numbers:  $\chi = 2(h_{11} - h_{12})$ vs $h_{11} +
h_{12}$. This plot has diagonal striations that we understand as
originating from the series of CY$_3$ spaces in two-vector chains, and
horizontal bands at (roughly)  constant $h_{11} + h_{12}$ that result from
combinations of these chains. The complete-intersection CY$_3$
(CICY) spaces correspond to higher-level
algebraic constructions in our approach, and
results are in preparation \cite{AENV}.

\section{Basic Features of the Construction}

In the search for CY spaces, one may consider~\cite{weight} weighted
complex
projective spaces $CP^n(k_1, k_2, ..., k_{n+1})$, which are
characterized by $(n+1)$ quasihomogeneous coordinates
$z_1, z_2, ..., z_{n+1}$, with the identification:
\begin{equation}
(z_1, z_2, ..., z_{n+1}) \sim (\lambda^{k_1}.z_1, \lambda^{k_2}.z_2, ...,
\lambda^{k_{n+1}}.z_{n+1})
\label{quasihom}
\end{equation}
As is well known, the loci of quasihomogeneous polynomial equations
in such complex weighted projective spaces yield compact
submanifolds, as discussed in more detail in~\cite{AENV} and references
therein. Consider a general polynomial $\cal P$ of degree $d$:
\begin{equation}
{\cal P} \equiv \Sigma_{\vec \mu} \, c_{\vec \mu} x^{\vec \mu}
\label{polynomial}
\end{equation}
which is a linear combination of monomials
$x^{\vec \mu} \equiv x_1^{\mu_1} \, . \, x_2^{\mu_2}. ...
.x_{n+1}^{\mu_{n+1}}$
with the condition
\begin{equation}
{\vec \mu} \, . \, {\vec k} = d. 
\label{mukcondition}
\end{equation}
The key to our approach
is a systematic construction of all possible projective vectors
$\vec k$, proceeding from lower-dimensional complex spaces to
higher-dimensional ones.

We recall that Batyrev~\cite{Bat} found an elegant characterization of
CY manifolds in terms of the corresponding Newton polyhedra,
defined as the complex hulls of all the vectors $\vec \mu$
satisfying (\ref{mukcondition}), which provides a systematic
approach to duality~\cite{duality} and mirror symmetry~\cite{mirror}.
Defining the vector
${\vec \mu'} \equiv {\vec \mu} - (1,1, ... ,1)$, which obeys ${\vec
\mu'}.{\vec k} =0$ and is henceforward denoted without the prime
($\prime$),
we define the lattice $\Lambda$:
\begin{equation}
\Lambda \equiv \{ {\vec \mu}: {\vec \mu} \, . \, {\vec k} = 0 \},
\label{lattice}
\end{equation}
with basis vectors ${\vec e_i}$. Consider now the polyhedron
$\Delta$, defined as the complex hull of all the vectors $\vec \mu$
in $\Lambda$ which have all components $\mu_i \ge -1$. Batyrev
showed that, in order to describe a CY space, such a polyhedron
must be reflexive, i.e.,

$\bullet$ the vertices of the polyhedron should correspond to
vectors $\vec \mu$ with integer components, 

$\bullet$ there should be just one interior point, called the center, and

$\bullet$ the distance of any face of the polyhedron from the center
should be unity.

\noindent
He also showed that the mirror $\Delta^*$ of any reflexive
polyhedron is also reflexive, and hence defines the corresponding
mirror CY space.

Kreuzer and Skarke~\cite{KS} have enumerated the possible reflexive
polyhedra in four complex dimensions, and provided a scatter plot
of the Hodge numbers of the corresponding CY$_3$ spaces. Here we
reveal some of the algebraic structure of the web of CY$_3$ spaces,
based on relations between the underlying projective vectors $\vec k$.

This constructive approach starts~\cite{AENV} from 
basic projective vectors ${\vec k}_N$
in a lower dimension, and extends them to higher-dimensional
projective vectors: ${\vec k}_{N+1} \equiv (0, {\vec k}_N)$ and
similarly for the other $(N+1)$-dimensional vectors obtained by
adding the `zero' coordinate in all the $N$ other possible ways. 
In the case of K3 spaces, for example, we found~\cite{AENV} a total of 95
such basic projective vectors in four dimensions. These may be
extended to 100 different types of projective vector in five
dimensions, yielding 10~270 distinct vectors when permutations
are taken into account, which we may then use to contruct
CY$_3$ spaces.

To do so, one must combine the basic vectors in all the possible ways
that yield distinct reflexive polynomials. In the case of K3 spaces, we
found the 95 basic projective vectors by considering 22 two-vector
`chains' (discrete linear combinations of pairs of simple extensions of
three-dimensional vectors) and 4 three-vector `chains', and there was one
`odd vector out' that we found using duality. There was considerable
overlap between the different chains: most of the 95 K3 projective
vectors appeared in more than one chain, and some appeared in as many as
five two-vector chains and three three-vector chains. The topological
properties of different spaces in the same chain display systematic
relations, as discussed in~\cite{AENV} and again later in this paper. 

As we mentioned in~\cite{AENV}, the 10~270 distinct basic projective
vectors in five dimensions may combined in 4~242 two-vector chains,
259 three-vector chains, 6 four-vector chains and one five-vector
chain. The challenge in this approach is to generate all the
two- (three-, four-, five-) vector combinations that correspond to
distinct reflexive polynomials, and hence CY$_3$ spaces. In the next
Section we outline the results from this construction.

Our method displays immediately interesting fibrations~\cite{KLM,KV,AKMS}.
We recall that the following are equivalent necessary and sufficient
conditions for a CY$_n$ space to have as a fibration a
CY$_{n-k}$ space~\cite{CPR}:  (a) there is a projection operator $\Pi:
\Lambda
\rightarrow \Lambda_{n-k}$, where $\Lambda_{n-k}$ is an
$(n-k)$-dimensional sublattice with $\Pi(\Delta)$ a reflexive polyhedron,
or (b)  there is a plane of the dual lattice $\Lambda^*$ through the
origin whose intersection with the dual polyhedron $\Delta^*$ is an
$(n-k)$-dimensional reflexive polyhedron.  These conditions are easy to
study in our approach, and, as we discuss in the next Section, our two-,
three- and four-vector chains naturally correspond to K3, elliptic and
reflexive line-segment {base-sections}. 

\section{Chains of CY$_3$ Spaces}

\subsection{Two-Vector Chains}

As already mentioned, there are 4~242 such two-vector chains $m_1
{\vec k}^{(1,ext)} + m_2 {\vec k}^{(2,ext)}$, where $m_1$ and $m_2$ are
integers. We call the vector with the lowest values of $m_1=1$ and $m_2=1$
the eldest vector, and that with the highest value of $m_1$ and $m_2$ the
youngest vector. Since the maximum values of $m_1$ and $m_2$ are
restricted by the dimensions of the extended vectors: $m_1 \leq {\rm dim}
[{\vec k}^{(2,ext)}]$ and $m_2 \leq {\rm dim} [{\vec k}^{(1,ext)}]$, it is
possible to find in a straightforward but laborious manner all projective
vectors in all 4~242 two-vector chains. However, the analogous process
becomes prohibitively lengthy in the case of the 259 three-vector chains,
so it is desirable to accelerate the process of determining which pairs
$m_1$ and $m_2$ yield projective vectors that correspond to reflexive
polyhedra and hence $CY_3$ spaces. To this end, we have developed an
`expansion' procedure, which we illustrate in one of the simpler
two-vector cases, thereby illuminating some of the geometric features of
our construction.

\begin{figure}[htb]  
\vskip -3cm
\hspace*{-.35in}
\begin{minipage}{8in}
\epsfig{figure=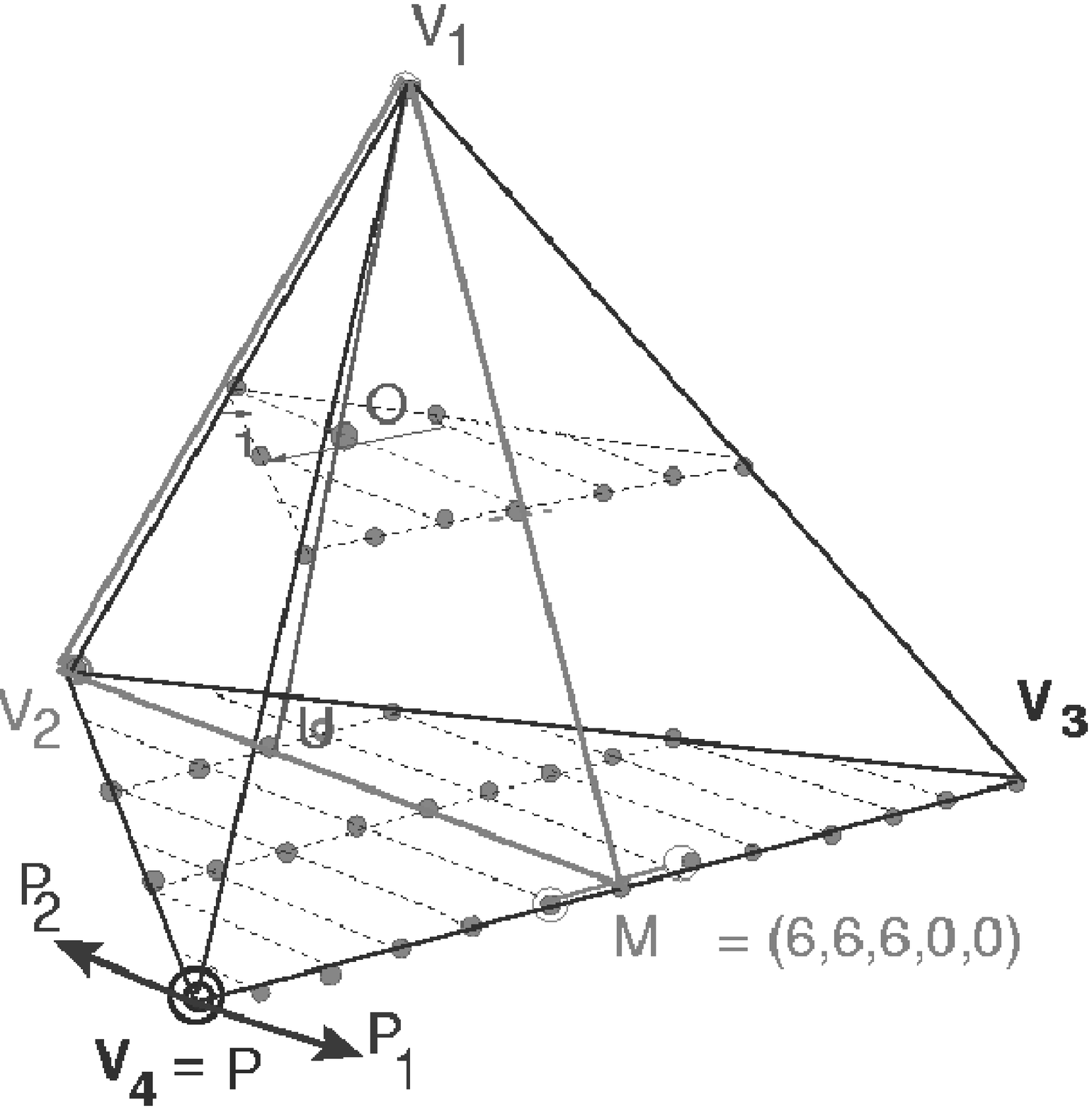,height=4.in}
\hspace*{-.65in}\epsfig{file=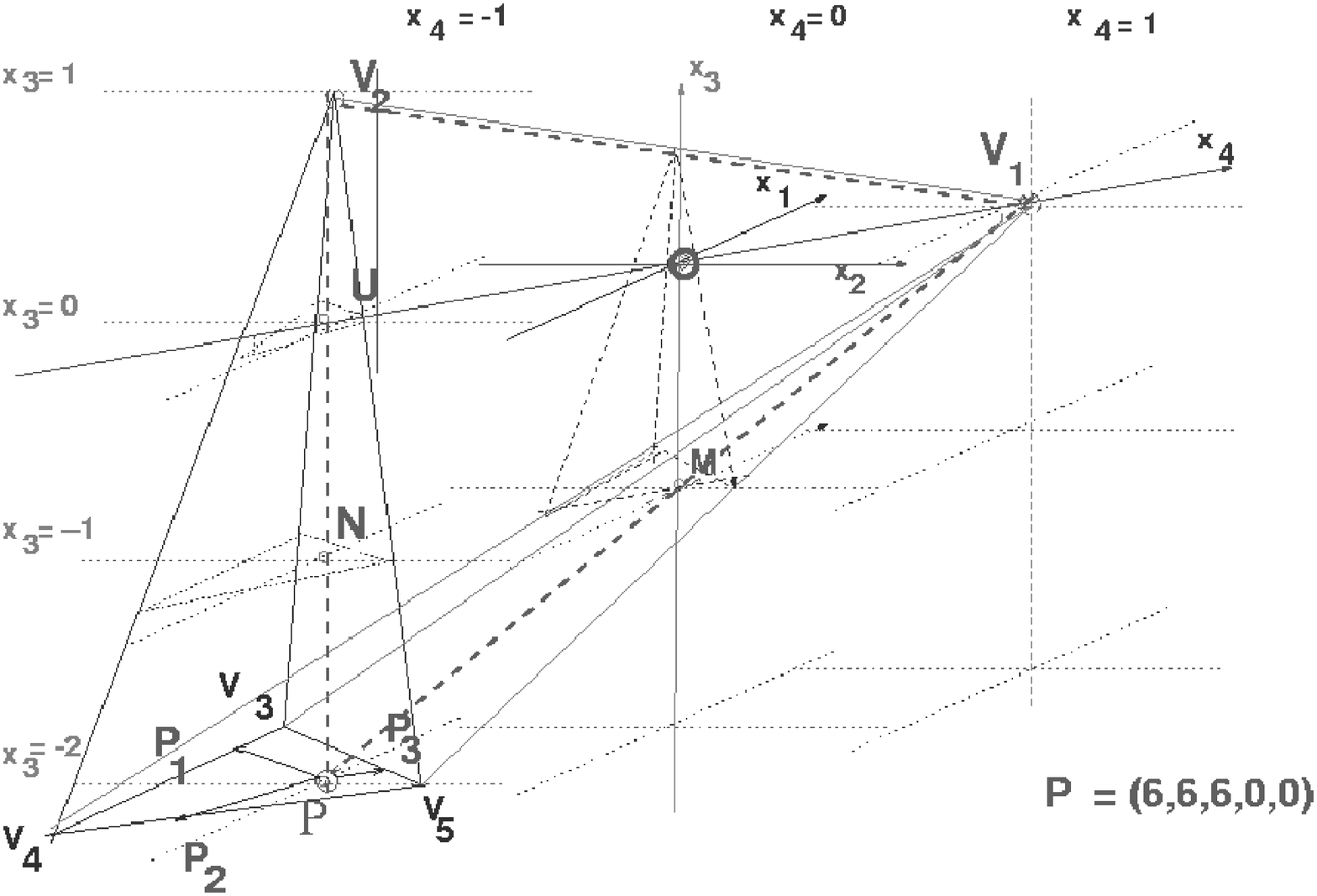,height=4.in} \hfill
\end{minipage}
\vskip -0.5cm
\caption{\it (a) The `expansion' method for CY$_3$ spaces in the
two-vector chain: $m(1,0,1,4,6) + n(0,1,1,4,6)$, which intersect in the
K3 space generated by $k_4 = (1,1,4,6)$~\cite{AENV}, and
(b) the `double-expansion' method for the
three-vector chain $m(1,0,0,2,3) + n(0,1,0,2,3) + l(0,0,1,2,3)$
with an elliptic Weierstrass fibre~\cite{AENV}, indicated by the dashed
triangle. The double-expansion point for this chain is $P=(6,6,6,0,0)$.}
\label{fig:vexp}
\end{figure}

Consider the chain generated by the extended vectors ${\vec k}^{(1,ext)}
\equiv (1,0,1,4,6)$, ${\vec k}^{(2,ext)} \equiv (0,1,1,4,6)$, namely
$m(1,0,1,4,6) +n(0,1,1,4,6) = (m,n,m+n,4m+4n,6m+6n)[12m+12n]$, where the
final number $[N]$ is just the sum of the vector components. All the
CY$_3$ spaces generated by this chain contain~\cite{CPR} a K3 fiber
associated with
the four-dimensional projective vector ${\vec k}_4 \equiv (1,1,4,6)$,
which defines the three-dimensional polyhedron of monomials $\vec
\mu$ shown in
Fig.~\ref{fig:vexp}(a). In order to
construct the full list of the projective vectors
$\vec{k}_5$ in each such two-vector chain generated
in our algebraic approach, we `expand' around a suitable vertex $P$ of the
corresponding three-dimensional
polyhedron, such as the one shown in Fig.~\ref{fig:vexp}(a). This is
done by finding all the possible pairs of
positive integer points $P_1, P_2$:
\begin{eqnarray}
P\, = \, \frac{1}{2}(P_1\,+\,P_2)
\end{eqnarray}
The set of possible pairs $P_1 $ and $P_2$ correspond to the set
of all the possible integer pairs $m_1$ and $m_2$ that determine the
two-vector chain. In this way,
a reflexive polyhedron of dimension $D=3$ is expanded
into reflexive polyhedra of dimension $D+1=4$.
In terms of the expansion point $P$ and the other
invariant vertices $V_1,V_2,V_3$, 
the allowed new four-dimensional reflexive polyhedra 
can be described by the projective vectors $\vec{k}_5$
that satisfy the following chain equations:
\begin{eqnarray}
{\vec k}_5\, . \, P_1\,&=&\,{\vec k}_5\, . \, P_2\,=\,d,\nonumber\\
{\vec k}_5\, . \, V_1\,&=&\,{\vec k}_5\, . \, V_2\,=\,{\vec k}_5\, .
\, V_3\,=\,d 
\end{eqnarray}
This procedure is illustrated in Fig.~\ref{fig:vexp}(a)
for the K3-fibre chain: $m(1,0,1,4,6) +n(0,1,1,4,6)=
(m,n,m+n,4m+4n,6m+6n)[12m+12n]$. 
We choose as the point of expansion
$P \equiv V_4 \equiv (12,12,0,0,0)$, and other three other vertices
$V_1 \equiv (0,0,0,0,2), V_2 \equiv (0,0,0,3,0), V_3 \equiv (0,0,12,0,0)$
are left invariant.

\begin{flushleft}
\begin{figure}[htb]
\epsfig{figure=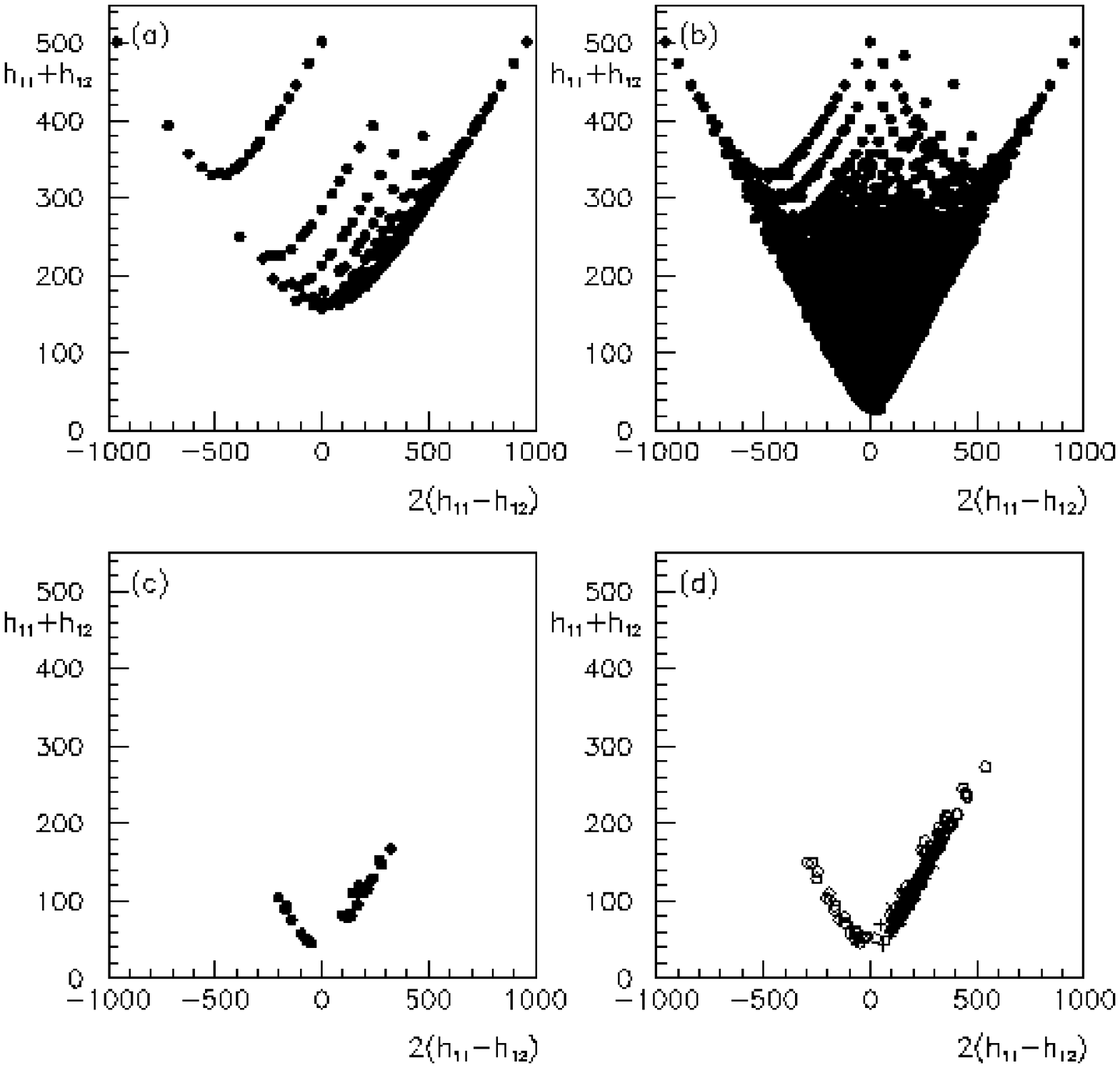,height=5.in}
\caption{\it Scatter plots of the Hodge numbers (a) for all the 542
CY$_3$
spaces in one of the 4~242 two-vector chains, namely
$m (0, 1, 6, 14, 21) + n (1, 0, 6, 14, 21)$, (b) for all the CY$_3$ spaces
constructed via two-vector
chains, (c) for the additional CY$_3$ spaces, not
obtainable from any of the two-vector chains, that are found in the
three-vector chain $m (0, 0, 1, 1, 1) + n (0 , 1, 0, 1, 2) + l (
1, 0, 0, 1, 2)$, and (d) for all the additional CY$_3$
spaces constructed via chains obtained from three (circles)
four (crosses) and five vectors (plus signs).}
\label{fig:chains}
\end{figure}
\end{flushleft}

The topological invariants of the CY$_3$ spaces associated with different
members of a chain are easily evaluated~\cite{AENV}, and we display
in Fig.~\ref{fig:chains}(a) the Hodge numbers
for the 542 CY$_3$ spaces in the two-vector chain:
$m(0,1,6,14,21) + n(1,0,6,14,21)$. 
The vectors in this chain have between them 206
distinct values of the Hodge numbers $h_{11}, h_{12}$, which are
visible in Fig.~\ref{fig:chains}(a). There may be equivalences
between the CY$_3$ spaces defined by different projective vectors ${\vec
k}_5$ and their corresponding reflexive polyhedra, generated by
modular transformations on the coordinates. However, these cannot change
the number of points, vertices or edges in either the polyhedron, $(N, V,
E)$, or
its mirror, $(N^*, V^*, E^*)$. The 542 CY$_3$ spaces in this two-vector
chain have between them 222 distinct values of the sextuple
$(N, N^*, V, V^*, h_{11}, h_{12})$. There are many other
topological invariants that may discriminate further between the
542 CY$_3$ spaces in this two-vector chain, but a complete exploration
of them goes beyond the scope of this paper.

The striations visible in
Fig.~\ref{fig:chains}(a) correspond to varying values of $m$ (or $n$) 
for some fixed value of $n$ (or $m$) in the chain 
$m (0, 1, 6, 14, 21) + n (1, 0, 6, 14, 21)$. These features reappear 
in Fig.~\ref{fig:chains}(b), which shows the full set of pairs of Hodge
numbers
found using all the 4~242 different two-vector chains, which
contain a total of 174~767 distinct projective vectors ${\vec k}_5$. 
By construction, all the CY$_3$ spaces found in
these two-vector chains have K3 {base-sections}~\cite{KLM,KV,AKMS,CPR}, 
as in the
example discussed above. It is interesting to note
that this scatter plot contains 10~827 distinct values of the pairs
$h_{11}, h_{12}$. We note that it contains
structures reflecting the striations seen in the simple example of
Fig.~\ref{fig:chains}(a). This explains the origin of similar structures
visible in the
scatter plot shown by Kreuzer and Skarke~\cite{KS}, which contains a total
of 30~108 different points. 

\begin{flushleft}
\begin{figure}[htb]
\epsfig{figure=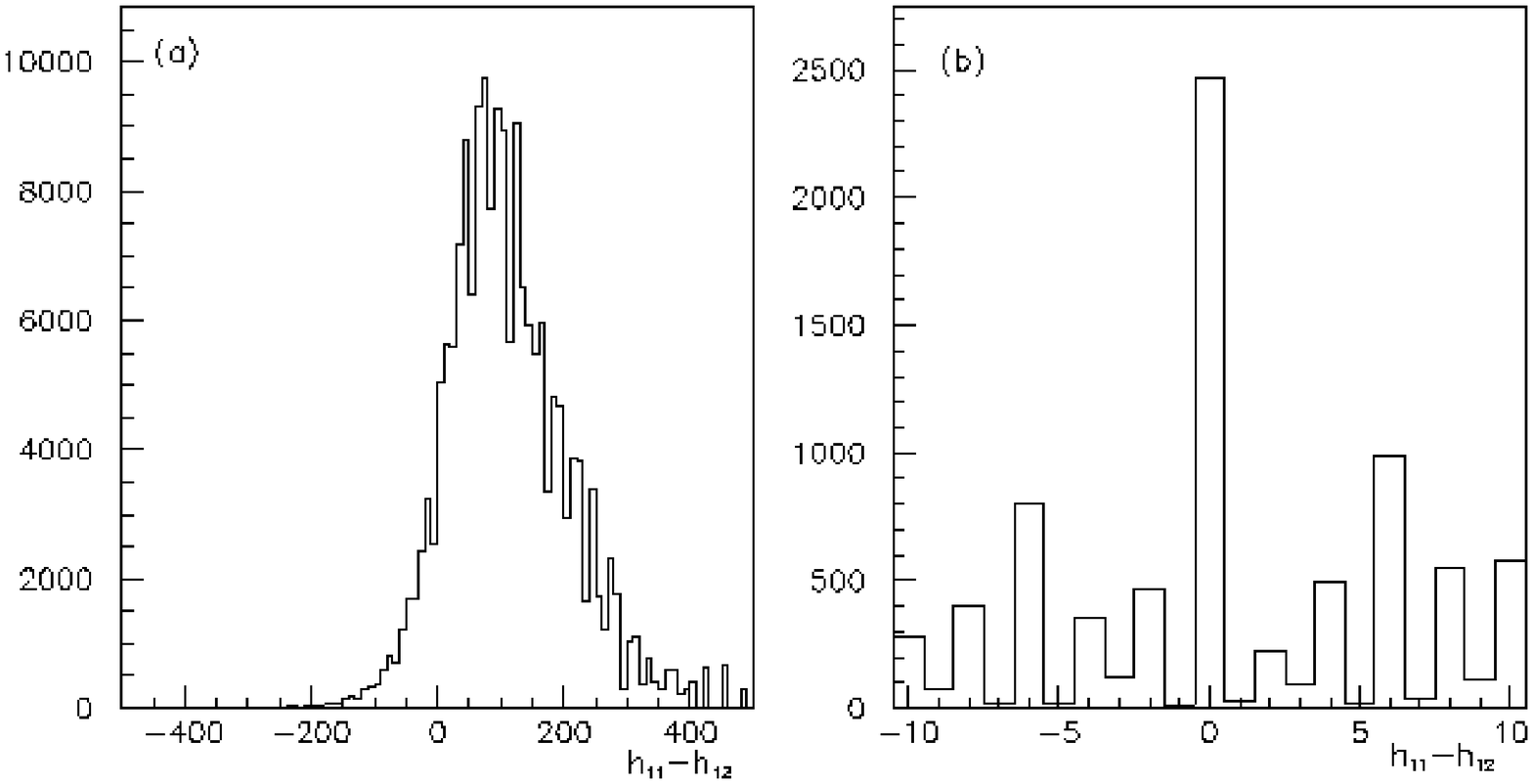,height=2.5in}
\caption{\it Histograms of the CY$_3$  
spaces constructed via two-vector chains, as functions of 
$h_{11} - h_{12}$ and hence the number of
generations $N_g = |h_{11} - h_{12}|$, (a) for all the CY$_3$ spaces
found, and (b) for the CY$_3$ spaces and mirrors with $N_g \le 10$.}
\label{fig:histograms}
\end{figure}
\end{flushleft}

Fig.~\ref{fig:histograms} displays histograms of CY$_3$ spaces with
different values of $h_{11} - h_{12}$ and hence the number of generations
$N_g = |h_{11} - h_{12}|$. The CY$_3$
manifolds of most
phenomenological interest are those with $h_{11} - h_{12} = 3$ and their
mirror
manifolds with $h_{11} - h_{12} = -3$. The larger-scale histogram of
Fig.~\ref{fig:histograms}(a) is blown up in Fig.~\ref{fig:histograms}(b),
for a close-up view concentrating on small values of $N_g$. As already
mentioned in the Introduction, we see a total of 94 
spaces with $h_{11} - h_{12} =
3$, and a total of 118 with $h_{11} - h_{12} = - 3$, whose mirrors have
$N_g = 3$.  The corresponding projective
vectors, Hodge numbers, and relevant
properties of the associated reflexive polyhedra are listed in the Tables.
Among these, there is just one
mirror pair, leaving us with 211 three-generation CY$_3$ spaces.  Between
them, these spaces have 99 + 81 - 1 = 179 distinct values of the octuple
$(N, N^*, V, V^*, E, E^*, h_{11}, h_{12})$. Thus, we have a total of at
least
179 distinct three-generation manifolds at our disposal, all of which
possess, by construction, K3 sections. We
have not explored systematically the gauge groups that may be associated
with their singularities, which would be straightforward in principle. As
can be inferred from Figs.~\ref{fig:chains}(a)  and
\ref{fig:histograms}(b), these three-generation spaces are frequently
related to manifolds that are `nearby' in chains, possibly with `similar'
values of $N_g$. A systematic study of these features might provide
interesting insights into transitions between different CY$_3$ vacua, but
goes beyond the scope of this paper. 

\begin{center}
  \begin{table}[!ht]
\caption{{\it Listing of the 94 five-dimensional projective vectors
${\vec k}_5$ defining three-generation
CY$_3$ spaces with K3 {sections}, including their Hodge numbers and
quantities associated
with the corresponding reflexive polyhedra.}}
  {\tiny
 
  \begin{tabular} {|c||c||c|c||c|c||c|c|||c||c||c|c||c|c||c|c|}
  \hline
$ \aleph $&$ {\vec k}_5 $&$ N $&$ N^* $&$ h_{12} $&$ h_{11} $&$V $&$ V^*
$&$ \aleph $&$ {\vec k}_5
$&$ N $&$ N^* $&$ h_{12} $&$ h_{11} $&$V $&$ V^* $\\
  \hline
$   1$&$(  2,  3,  8, 11, 17) $&$ 62$&$ 57$&$ 45$&$ 48$&$ 16$&$ 13$&  
$   2$&$(  2,  5,  9, 10, 11) $&$ 37$&$ 34$&$ 27$&$ 30$&$ 13$&$ 10$\\
  \hline
$   3$&$(  2,  5, 14, 21, 33) $&$ 65$&$ 59$&$ 47$&$ 50$&$ 12$&$ 11$&
$   4$&$(  3,  4,  5,  8, 13) $&$ 38$&$ 36$&$ 27$&$ 30$&$ 14$&$ 12$\\
  \hline
$   5$&$(  3,  4, 10, 15, 17) $&$ 37$&$ 38$&$ 27$&$ 30$&$ 15$&$ 13$&
$   6$&$(  3,  5, 15, 16, 24) $&$ 39$&$ 40$&$ 31$&$ 34$&$  9$&$  6$\\ 
  \hline
$   7$&$(  3,  5, 16, 24, 45) $&$ 60$&$ 51$&$ 43$&$ 46$&$ 14$&$ 11$&
$   8$&$(  4,  5,  6, 15, 21) $&$ 37$&$ 35$&$ 26$&$ 29$&$ 15$&$ 13$\\
  \hline
$   9$&$(  4,  5,  7,  8,  9) $&$ 24$&$ 25$&$ 17$&$ 20$&$ 13$&$ 13$&
$  10$&$(  4,  5,  7,  9, 10) $&$ 24$&$ 21$&$ 17$&$ 20$&$ 11$&$ 11$\\   
  \hline
$  11$&$(  4,  6,  7,  9, 17) $&$ 26$&$ 30$&$ 19$&$ 22$&$ 14$&$ 17$&
$  12$&$(  3,  7,  8, 10, 25) $&$ 37$&$ 36$&$ 26$&$ 29$&$ 15$&$ 16$\\
  \hline
$  13$&$(  3,  5,  6,  8, 17) $&$ 37$&$ 31$&$ 25$&$ 28$&$ 13$&$ 13$&
$  14$&$(  2,  7,  9, 10, 13) $&$ 34$&$ 34$&$ 25$&$ 28$&$ 14$&$ 15$\\
  \hline
$  15$&$(  2,  8,  9, 11, 21) $&$ 40$&$ 39$&$ 29$&$ 32$&$ 14$&$ 14$&
$  16$&$(  3,  4,  6, 13, 13) $&$ 38$&$ 17$&$ 26$&$ 29$&$  9$&$  8$\\
  \hline
$  17$&$(  5,  8,  9, 11, 12) $&$ 18$&$ 22$&$ 13$&$ 16$&$ 11$&$ 13$&
$  18$&$(  2,  7,  8, 11, 17) $&$ 38$&$ 39$&$ 28$&$ 31$&$ 16$&$ 17$\\
  \hline
$  19$&$(  4,  5,  7, 10, 13) $&$ 25$&$ 26$&$ 18$&$ 21$&$ 12$&$ 12$&
$  20$&$(  2,  6,  9, 17, 17) $&$ 43$&$ 22$&$ 31$&$ 34$&$ 10$&$  8$\\
  \hline
$  21$&$(  2,  8,  9, 21, 23) $&$ 44$&$ 47$&$ 33$&$ 36$&$ 11$&$ 10$&
$  22$&$(  2,  7, 12, 15, 21) $&$ 40$&$ 36$&$ 29$&$ 32$&$ 13$&$ 11$\\
  \hline
$  23$&$(  2,  8,  9, 27, 37) $&$ 64$&$ 62$&$ 46$&$ 49$&$ 10$&$ 10$&
$  24$&$(  2,  8, 21, 41, 51) $&$ 62$&$ 71$&$ 47$&$ 50$&$ 11$&$ 12$\\
  \hline
$  25$&$(  2, 11, 12, 39, 53) $&$ 66$&$ 71$&$ 48$&$ 51$&$ 12$&$ 12$&
$  26$&$(  5,  6,  9, 14, 17) $&$ 22$&$ 23$&$ 15$&$ 18$&$ 11$&$ 12$\\
  \hline
$  27$&$(  3,  7,  8,  9, 10) $&$ 25$&$ 26$&$ 18$&$ 21$&$ 14$&$ 14$&
$  28$&$(  4,  5,  6, 21, 27) $&$ 46$&$ 48$&$ 33$&$ 36$&$ 14$&$ 13$\\
  \hline
$  29$&$(  4,  6, 15, 35, 45) $&$ 44$&$ 47$&$ 34$&$ 37$&$ 10$&$  9$&
$  30$&$(  4,  5,  8, 10, 17) $&$ 29$&$ 27$&$ 22$&$ 25$&$  8$&$  7$\\
  \hline
$  31$&$(  2,  5, 11, 20, 27) $&$ 59$&$ 49$&$ 42$&$ 45$&$ 11$&$ 12$&
$  32$&$(  2,  5, 17, 26, 33) $&$ 64$&$ 66$&$ 47$&$ 50$&$ 17$&$ 17$\\
  \hline
$  33$&$(  3,  4, 17, 27, 30) $&$ 54$&$ 55$&$ 41$&$ 44$&$ 14$&$ 14$&
$  34$&$(  3,  4, 21, 32, 39) $&$ 64$&$ 70$&$ 50$&$ 53$&$ 12$&$ 13$\\
  \hline
$  35$&$(  5,  6,  8, 33, 47) $&$ 50$&$ 53$&$ 35$&$ 38$&$ 15$&$ 15$&
$  36$&$(  3,  5, 14, 20, 21) $&$ 36$&$ 32$&$ 25$&$ 28$&$  9$&$  9$\\
  \hline
$  37$&$(  4,  5, 11, 29, 38) $&$ 47$&$ 45$&$ 33$&$ 36$&$ 12$&$ 12$&
$  38$&$(  5,  6,  7, 29, 40) $&$ 47$&$ 45$&$ 32$&$ 35$&$ 11$&$ 11$\\
  \hline
$  39$&$(  3,  6, 10, 19, 35) $&$ 46$&$ 45$&$ 33$&$ 36$&$ 11$&$ 13$&
$  40$&$(  2,  5, 11, 20, 33) $&$ 68$&$ 57$&$ 48$&$ 51$&$ 12$&$ 11$\\
  \hline
%
$  41$&$(  2,  5, 13, 22, 37) $&$ 72$&$ 73$&$ 51$&$ 54$&$ 14$&$ 15$&
$  42$&$(  2,  5, 17, 26, 45) $&$ 80$&$ 79$&$ 57$&$ 60$&$ 12$&$ 13$\\
  \hline
$  43$&$(  2,  5, 23, 32, 57) $&$ 94$&$ 96$&$ 67$&$ 70$&$ 13$&$ 14$&
$  44$&$(  4,  5,  5, 18, 22) $&$ 43$&$ 34$&$ 29$&$ 32$&$  9$&$  7$\\
  \hline
$  45$&$(  3,  9, 10, 11, 24) $&$ 33$&$ 38$&$ 29$&$ 32$&$ 13$&$ 14$&
$  46$&$(  3,  9, 14, 15, 16) $&$ 29$&$ 44$&$ 33$&$ 36$&$ 11$&$ 12$\\
  \hline
$  47$&$(  5,  7,  9, 15, 36) $&$ 32$&$ 30$&$ 21$&$ 24$&$  9$&$  8$&
$  48$&$(  4,  5,  7, 18, 20) $&$ 34$&$ 35$&$ 24$&$ 27$&$ 11$&$  9$\\
  \hline
$  49$&$(  5,  6,  9, 10, 21) $&$ 27$&$ 30$&$ 20$&$ 23$&$ 13$&$ 12$&
$  50$&$(  1, 10, 12, 13, 15) $&$ 58$&$ 55$&$ 43$&$ 46$&$ 10$&$ 11$\\
  \hline
$  51$&$(  1, 11, 13, 14, 16) $&$ 58$&$ 55$&$ 43$&$ 46$&$ 10$&$ 11$&
$  52$&$(  5,  6,  9, 12, 13) $&$ 21$&$ 26$&$ 17$&$ 20$&$  9$&$ 11$\\
  \hline
$  53$&$(  4,  5,  7,  8, 11) $&$ 25$&$ 27$&$ 18$&$ 21$&$ 12$&$ 11$&
$  54$&$(  1,  8, 13, 14, 17) $&$ 63$&$ 60$&$ 47$&$ 50$&$ 10$&$ 10$\\
  \hline
$  55$&$(  1, 14, 23, 24, 31) $&$ 63$&$ 60$&$ 47$&$ 50$&$ 10$&$ 10$&
$  56$&$(  1, 11, 18, 19, 24) $&$ 63$&$ 60$&$ 47$&$ 50$&$ 10$&$ 10$\\
  \hline
$  57$&$(  4,  5, 10, 11, 19) $&$ 29$&$ 31$&$ 20$&$ 23$&$ 12$&$ 12$&
$  58$&$(  4,  4,  9, 11, 17) $&$ 32$&$ 34$&$ 22$&$ 25$&$  8$&$  8$\\
  \hline
$  59$&$(  4,  5,  8, 11, 17) $&$ 28$&$ 32$&$ 20$&$ 23$&$ 11$&$ 12$&
$  60$&$(  1,  5, 11, 16, 18) $&$ 84$&$ 78$&$ 62$&$ 65$&$ 13$&$ 12$\\
  \hline
$  61$&$(  1,  8, 18, 27, 29) $&$ 84$&$ 78$&$ 62$&$ 65$&$ 13$&$ 12$&
$  62$&$(  1,  9, 20, 30, 33) $&$ 84$&$ 78$&$ 62$&$ 65$&$ 13$&$ 12$\\
  \hline
$  63$&$(  1,  8, 10, 21, 23) $&$ 81$&$ 81$&$ 60$&$ 63$&$ 12$&$ 12$&
$  64$&$(  5,  6,  7,  8, 13) $&$ 22$&$ 24$&$ 15$&$ 18$&$ 14$&$ 14$\\
  \hline
$  65$&$(  1,  8, 11, 20, 23) $&$ 77$&$ 75$&$ 57$&$ 60$&$ 14$&$ 14$&
$  66$&$(  3, 12, 14, 21, 25) $&$ 29$&$ 47$&$ 33$&$ 36$&$ 10$&$ 11$\\
  \hline
$  67$&$(  2,  9, 11, 24, 35) $&$ 55$&$ 49$&$ 37$&$ 40$&$ 11$&$ 11$&
$  68$&$(  3,  5,  8, 22, 33) $&$ 58$&$ 54$&$ 40$&$ 43$&$ 13$&$ 14$\\
  \hline
$  69$&$(  1,  8, 10, 21, 31) $&$ 97$&$ 88$&$ 70$&$ 73$&$ 13$&$ 11$&
$  70$&$(  4,  5,  7, 20, 31) $&$ 46$&$ 46$&$ 32$&$ 35$&$ 13$&$ 13$\\
  \hline
$  71$&$(  3,  4, 15, 38, 57) $&$ 98$&$ 99$&$ 74$&$ 77$&$ 12$&$ 11$&
$  72$&$(  3,  8, 30, 79,117) $&$ 98$&$ 99$&$ 74$&$ 77$&$ 12$&$ 11$\\
  \hline
$  73$&$(  2,  9, 12, 23, 37) $&$ 51$&$ 48$&$ 37$&$ 40$&$ 12$&$ 14$&
$  74$&$(  3,  4, 13, 24, 41) $&$ 67$&$ 68$&$ 47$&$ 50$&$ 13$&$ 13$\\
  \hline
$  75$&$(  4,  5,  7, 23, 34) $&$ 51$&$ 53$&$ 36$&$ 39$&$ 14$&$ 15$&
$  76$&$(  5,  6, 14, 45, 65) $&$ 56$&$ 60$&$ 42$&$ 45$&$ 12$&$ 11$\\
  \hline
$  77$&$(  4,  7, 13, 41, 58) $&$ 53$&$ 53$&$ 37$&$ 40$&$ 13$&$ 13$&
$  78$&$(  5,  8, 12, 15, 35) $&$ 30$&$ 40$&$ 27$&$ 30$&$ 11$&$ 11$\\
  \hline
$  79$&$(  2,  7, 15, 18, 21) $&$ 40$&$ 35$&$ 29$&$ 32$&$ 11$&$ 10$&
$  80$&$(  2,  9, 19, 24, 27) $&$ 40$&$ 35$&$ 29$&$ 32$&$ 11$&$ 10$\\
  \hline
$  81$&$(  3,  7, 15, 18, 20) $&$ 32$&$ 40$&$ 29$&$ 32$&$  8$&$ 11$&
$  82$&$(  3,  8, 18, 21, 25) $&$ 32$&$ 40$&$ 29$&$ 32$&$  8$&$ 11$\\
  \hline
$  83$&$(  1, 15, 22, 28, 33) $&$ 63$&$ 65$&$ 47$&$ 50$&$ 10$&$ 10$&
$  84$&$(  1, 18, 26, 33, 39) $&$ 63$&$ 65$&$ 47$&$ 50$&$ 10$&$ 10$\\
  \hline
$  85$&$(  1, 21, 30, 38, 45) $&$ 63$&$ 65$&$ 47$&$ 50$&$ 10$&$ 10$&
$  86$&$(  1, 11, 16, 20, 23) $&$ 63$&$ 65$&$ 47$&$ 50$&$ 10$&$ 10$\\
  \hline
$  87$&$(  1, 13, 19, 24, 28) $&$ 63$&$ 65$&$ 47$&$ 50$&$ 10$&$ 10$&
$  88$&$(  1, 16, 23, 29, 34) $&$ 63$&$ 65$&$ 47$&$ 50$&$ 10$&$ 10$\\
  \hline
$  89$&$(  7, 10, 11, 14, 35) $&$ 26$&$ 40$&$ 27$&$ 30$&$  9$&$ 11$&
$  90$&$(  3,  8, 21, 43, 54) $&$ 52$&$ 59$&$ 42$&$ 45$&$  9$&$ 10$\\
  \hline
$  91$&$(  4,  5, 26, 65, 95) $&$ 91$&$ 95$&$ 67$&$ 70$&$ 11$&$ 11$&
$  92$&$(  2,  9, 26, 65, 93) $&$100$&$ 99$&$ 71$&$ 74$&$ 10$&$ 11$\\
  \hline
$  93$&$(  2,  9, 30, 73,105) $&$109$&$118$&$ 80$&$ 83$&$ 11$&$ 11$&
$  94$&$( 10, 12, 13, 15, 25) $&$ 16$&$ 28$&$ 17$&$ 20$&$  8$&$  9$\\
  \hline
 \end{tabular}
}
 \end{table}
\end{center}

\begin{center}
  \begin{table}[!ht]
\caption{{\it Listing of the 118 five-dimensional projective vectors
${\vec k}_5$ defining mirrors of
three-generation
CY$_3$ spaces with K3 {sections}, including their Hodge numbers and
quantities associated
with the corresponding reflexive polyhedra.}}
  {\tiny
\begin{tabular} {|c||c||c|c||c|c||c|c|||c||c||c|c||c|c||c|c|}
  \hline
$ \aleph $&$ {\vec k}_5  $&$ N $&$ N^* $&$ h_{12} $&$ h_{11} $&$V $&$ V^*
$&$ \aleph $&$ {\vec k}_5
$&$ N $&$ N^* $&$ h_{12} $&$ h_{11} $&$V $&$ V^* $\\
  \hline
$   1$&$(  2,  5,  6,  7,  7) $&$ 36$&$ 22$&$ 26$&$ 23$&$ 14$&$ 10$&
$   2$&$(  2,  5,  5,  9, 11) $&$ 42$&$ 26$&$ 30$&$ 27$&$  9$&$  7$\\
  \hline
$   3$&$(  2,  5, 13, 14, 21) $&$ 47$&$ 42$&$ 35$&$ 32$&$ 15$&$ 15$&
$   4$&$(  2,  5, 14, 21, 21) $&$ 53$&$ 26$&$ 38$&$ 35$&$ 12$&$  9$\\
  \hline
$   5$&$(  2,  6,  7,  7, 15) $&$ 42$&$ 24$&$ 30$&$ 27$&$ 10$&$  8$&
$   6$&$(  2,  6,  7, 15, 15) $&$ 44$&$ 22$&$ 32$&$ 29$&$ 13$&$ 10$\\
  \hline
$   7$&$(  3,  4,  7,  7, 10) $&$ 35$&$ 25$&$ 23$&$ 20$&$ 14$&$ 10$&
$   8$&$(  3,  4,  7, 14, 17) $&$ 43$&$ 32$&$ 30$&$ 27$&$ 13$&$ 11$\\
  \hline
$   9$&$(  3,  4,  9, 11, 18) $&$ 41$&$ 33$&$ 33$&$ 30$&$ 11$&$ 11$&
$  10$&$(  3,  4, 11, 18, 33) $&$ 60$&$ 42$&$ 43$&$ 40$&$ 13$&$ 10$\\
  \hline
$  11$&$(  3,  4, 14, 21, 21) $&$ 47$&$ 25$&$ 35$&$ 32$&$ 10$&$  8$&
$  12$&$(  4,  5,  6,  7, 17) $&$ 32$&$ 27$&$ 23$&$ 20$&$ 15$&$ 15$\\
  \hline
$  13$&$(  3,  5,  6,  8, 11) $&$ 31$&$ 19$&$ 21$&$ 18$&$ 10$&$ 10$&
$  14$&$(  3,  3,  5,  8, 14) $&$ 47$&$ 26$&$ 31$&$ 28$&$ 10$&$  9$\\
  \hline
$  15$&$(  3,  4,  6,  7, 13) $&$ 36$&$ 22$&$ 25$&$ 22$&$ 13$&$ 12$&
$  16$&$(  3,  5,  8,  9, 20) $&$ 38$&$ 24$&$ 26$&$ 23$&$  9$&$  9$\\
  \hline
$  17$&$(  3,  5,  8, 14, 15) $&$ 33$&$ 20$&$ 23$&$ 20$&$  9$&$  9$&
$  18$&$(  4,  6,  9, 11, 15) $&$ 25$&$ 22$&$ 22$&$ 19$&$ 10$&$  9$\\
  \hline
$  19$&$(  2,  6,  9, 13, 17) $&$ 40$&$ 35$&$ 30$&$ 27$&$ 13$&$ 15$&
$  20$&$(  2,  7,  8, 11, 13) $&$ 35$&$ 30$&$ 26$&$ 23$&$ 14$&$ 14$\\
  \hline
$  21$&$(  2,  8,  9, 13, 15) $&$ 35$&$ 31$&$ 26$&$ 23$&$ 12$&$ 13$&
$  22$&$(  4,  7,  9, 10, 15) $&$ 22$&$ 18$&$ 16$&$ 13$&$ 13$&$ 11$\\
  \hline
$  23$&$(  2,  7, 10, 11, 15) $&$ 35$&$ 30$&$ 26$&$ 23$&$ 14$&$ 13$&
$  24$&$(  2,  5, 11, 12, 15) $&$ 41$&$ 30$&$ 30$&$ 27$&$ 12$&$ 12$\\
  \hline
$  25$&$(  2,  4, 11, 21, 25) $&$ 64$&$ 66$&$ 49$&$ 46$&$ 11$&$ 11$&
$  26$&$(  2,  7, 10, 11, 23) $&$ 44$&$ 37$&$ 32$&$ 29$&$ 14$&$ 14$\\
  \hline
$  27$&$(  4,  4,  5,  5,  7) $&$ 29$&$ 14$&$ 20$&$ 17$&$  8$&$  7$&
$  28$&$(  2,  6, 11, 19, 27) $&$ 51$&$ 43$&$ 38$&$ 35$&$ 14$&$ 15$\\
  \hline
$  29$&$(  2,  6, 15, 23, 31) $&$ 53$&$ 46$&$ 40$&$ 37$&$ 12$&$ 14$&
$  30$&$(  2,  6,  7, 17, 19) $&$ 48$&$ 44$&$ 36$&$ 33$&$ 12$&$ 12$\\
  \hline
$  31$&$(  2,  7, 12, 13, 17) $&$ 36$&$ 32$&$ 27$&$ 24$&$ 13$&$ 13$&
$  32$&$(  2,  6, 13, 23, 31) $&$ 55$&$ 52$&$ 41$&$ 38$&$ 13$&$ 12$\\
  \hline
$  33$&$(  2,  6, 11, 25, 33) $&$ 62$&$ 55$&$ 46$&$ 43$&$ 10$&$ 11$&
$  34$&$(  2,  8, 15, 35, 45) $&$ 62$&$ 55$&$ 46$&$ 43$&$ 10$&$ 11$\\
  \hline
$  35$&$(  2,  7,  8, 17, 27) $&$ 52$&$ 42$&$ 38$&$ 35$&$ 15$&$ 16$&
$  36$&$(  2,  7,  8, 25, 35) $&$ 69$&$ 61$&$ 50$&$ 47$&$ 11$&$ 12$\\
  \hline
$  37$&$(  2,  8, 29, 49, 59) $&$ 66$&$ 71$&$ 51$&$ 48$&$ 10$&$ 11$&
$  38$&$(  2,  5, 10, 19, 31) $&$ 66$&$ 54$&$ 47$&$ 44$&$ 11$&$ 10$\\
  \hline
$  39$&$(  2, 10, 11, 35, 47) $&$ 65$&$ 58$&$ 47$&$ 44$&$ 10$&$ 11$&
$  40$&$(  2,  5, 12, 21, 35) $&$ 70$&$ 59$&$ 50$&$ 47$&$ 12$&$ 12$\\
  \hline
$  41$&$(  4,  5,  6, 21, 31) $&$ 51$&$ 45$&$ 36$&$ 33$&$ 14$&$ 12$&
$  42$&$(  2,  5,  9, 18, 25) $&$ 59$&$ 43$&$ 42$&$ 39$&$ 12$&$ 11$\\
  \hline
$  43$&$(  2,  5, 13, 22, 29) $&$ 60$&$ 55$&$ 44$&$ 41$&$ 14$&$ 16$&
$  44$&$(  2,  5, 15, 24, 31) $&$ 62$&$ 50$&$ 45$&$ 42$&$ 11$&$ 11$\\
  \hline
$  45$&$(  2,  5, 23, 32, 39) $&$ 72$&$ 66$&$ 53$&$ 50$&$ 16$&$ 17$&
$  46$&$(  2,  5, 25, 39, 46) $&$ 82$&$ 76$&$ 60$&$ 57$&$ 12$&$ 10$\\
  \hline
$  47$&$(  3,  5, 12, 15, 16) $&$ 35$&$ 34$&$ 31$&$ 28$&$  9$&$  8$&
$  48$&$(  3,  5, 10, 14, 16) $&$ 32$&$ 26$&$ 23$&$ 20$&$  8$&$  7$\\
  \hline
$  49$&$(  3,  4, 12, 17, 19) $&$ 40$&$ 34$&$ 29$&$ 26$&$ 11$&$ 12$&
$  50$&$(  3,  6,  8, 23, 37) $&$ 55$&$ 54$&$ 40$&$ 37$&$ 12$&$ 12$\\
  \hline
$  51$&$(  3,  4, 17, 31, 38) $&$ 62$&$ 48$&$ 45$&$ 42$&$ 11$&$ 12$&
$  52$&$(  3,  4, 33, 47, 54) $&$ 80$&$ 83$&$ 63$&$ 60$&$ 11$&$ 12$\\
  \hline
$  53$&$(  2,  9, 15, 20, 23) $&$ 37$&$ 32$&$ 27$&$ 24$&$ 12$&$ 11$&
$  54$&$(  3,  5, 45, 61, 69) $&$ 81$&$ 81$&$ 63$&$ 60$&$  7$&$  6$\\
  \hline
$  55$&$(  4,  5, 12, 15, 36) $&$ 45$&$ 27$&$ 36$&$ 33$&$  7$&$  6$&
$  56$&$(  2,  5, 14, 23, 39) $&$ 74$&$ 65$&$ 53$&$ 50$&$ 14$&$ 13$\\
  \hline
$  57$&$(  2,  5, 25, 34, 61) $&$ 98$&$ 88$&$ 70$&$ 67$&$ 11$&$ 11$&
$  58$&$(  3,  6,  8, 25, 33) $&$ 59$&$ 66$&$ 54$&$ 51$&$ 10$&$ 11$\\
  \hline
$  59$&$(  3,  9, 17, 22, 24) $&$ 33$&$ 44$&$ 38$&$ 35$&$ 10$&$ 12$&
$  60$&$(  3,  6, 10, 11, 27) $&$ 43$&$ 45$&$ 38$&$ 35$&$ 13$&$ 13$\\
  \hline
$  61$&$(  1,  4,  9, 13, 13) $&$ 85$&$ 32$&$ 62$&$ 59$&$  8$&$  7$&
$  62$&$(  1,  9, 11, 12, 12) $&$ 59$&$ 32$&$ 43$&$ 40$&$  9$&$  7$\\
  \hline
$  63$&$(  1,  8, 11, 13, 14) $&$ 60$&$ 52$&$ 45$&$ 42$&$ 12$&$ 12$&
$  64$&$(  1,  9, 12, 14, 15) $&$ 60$&$ 52$&$ 45$&$ 42$&$ 12$&$ 12$\\
  \hline
$  65$&$(  3, 12, 15, 16, 17) $&$ 28$&$ 35$&$ 32$&$ 29$&$ 10$&$ 11$&
$  66$&$(  1,  9, 13, 16, 22) $&$ 67$&$ 58$&$ 50$&$ 47$&$ 15$&$ 15$\\
  \hline
$  67$&$(  1,  7, 10, 12, 17) $&$ 67$&$ 58$&$ 50$&$ 47$&$ 15$&$ 15$&
$  68$&$(  1, 12, 17, 21, 30) $&$ 67$&$ 58$&$ 50$&$ 47$&$ 15$&$ 15$\\
  \hline
$  69$&$(  1, 10, 14, 17, 25) $&$ 67$&$ 58$&$ 50$&$ 47$&$ 15$&$ 15$&
$  70$&$(  1, 14, 20, 25, 35) $&$ 67$&$ 58$&$ 50$&$ 47$&$ 15$&$ 15$\\
  \hline
$  71$&$(  1,  9, 16, 19, 22) $&$ 67$&$ 60$&$ 50$&$ 47$&$  9$&$  9$&
$  72$&$(  1, 18, 32, 39, 45) $&$ 67$&$ 60$&$ 50$&$ 47$&$  9$&$  9$\\
  \hline
$  73$&$(  1,  8, 14, 17, 19) $&$ 67$&$ 60$&$ 50$&$ 47$&$  9$&$  9$&
$  74$&$(  1, 10, 18, 21, 25) $&$ 67$&$ 60$&$ 50$&$ 47$&$  9$&$  9$\\
  \hline
$  75$&$(  1, 13, 23, 28, 32) $&$ 67$&$ 60$&$ 50$&$ 47$&$  9$&$  9$&
$  76$&$(  1, 14, 25, 30, 35) $&$ 67$&$ 60$&$ 50$&$ 47$&$  9$&$  9$\\
  \hline
$  77$&$(  2,  5,  7,  7,  8) $&$ 36$&$ 23$&$ 26$&$ 23$&$ 13$&$  9$&
$  78$&$(  1,  8, 10, 11, 21) $&$ 73$&$ 56$&$ 53$&$ 50$&$ 13$&$ 13$\\
  \hline
$  79$&$(  3,  7, 11, 12, 30) $&$ 40$&$ 37$&$ 32$&$ 29$&$ 11$&$ 12$&
$  80$&$(  3,  8, 13, 15, 36) $&$ 40$&$ 37$&$ 32$&$ 29$&$ 11$&$ 12$\\
  \hline
$  81$&$(  2,  7, 10, 11, 19) $&$ 39$&$ 33$&$ 29$&$ 26$&$ 13$&$ 14$&
$  82$&$(  4,  5,  7,  8, 19) $&$ 33$&$ 27$&$ 23$&$ 20$&$ 13$&$ 11$\\
  \hline
$  83$&$(  5, 12, 13, 15, 20) $&$ 22$&$ 30$&$ 26$&$ 23$&$  8$&$ 11$&
$  84$&$(  1,  8, 10, 11, 27) $&$ 87$&$ 75$&$ 63$&$ 60$&$ 10$&$ 10$\\
  \hline
$  85$&$(  1,  9, 11, 12, 30) $&$ 87$&$ 75$&$ 63$&$ 60$&$ 10$&$ 10$&
$  86$&$(  3,  4,  5, 13, 14) $&$ 42$&$ 35$&$ 31$&$ 28$&$ 12$&$ 13$\\
  \hline
$  87$&$(  3,  5,  7,  8, 10) $&$ 29$&$ 24$&$ 21$&$ 18$&$ 11$&$ 12$&
$  88$&$(  3,  9, 12, 19, 20) $&$ 34$&$ 61$&$ 50$&$ 47$&$  7$&$  9$\\
  \hline
$  89$&$(  2,  5, 13, 18, 19) $&$ 47$&$ 39$&$ 35$&$ 32$&$ 14$&$ 13$&
$  90$&$(  3,  8, 21, 30, 31) $&$ 36$&$ 39$&$ 32$&$ 29$&$  9$&$ 10$\\
  \hline
$  91$&$(  3,  7, 18, 26, 27) $&$ 36$&$ 39$&$ 32$&$ 29$&$  9$&$ 10$&
$  92$&$(  2,  7,  7,  8, 12) $&$ 37$&$ 22$&$ 30$&$ 27$&$  8$&$  7$\\
  \hline
$  93$&$(  2,  9, 11, 16, 19) $&$ 35$&$ 31$&$ 26$&$ 23$&$ 12$&$ 12$&
$  94$&$(  4,  4,  7, 13, 15) $&$ 33$&$ 28$&$ 23$&$ 20$&$ 10$&$  9$\\
  \hline
$  95$&$(  1, 12, 16, 23, 29) $&$ 67$&$ 60$&$ 50$&$ 47$&$ 13$&$ 13$&
$  96$&$(  1, 10, 13, 19, 24) $&$ 67$&$ 60$&$ 50$&$ 47$&$ 13$&$ 13$\\
  \hline
$  97$&$(  1, 14, 19, 27, 34) $&$ 67$&$ 60$&$ 50$&$ 47$&$ 13$&$ 13$&
$  98$&$(  3,  5,  8, 21, 34) $&$ 60$&$ 49$&$ 41$&$ 38$&$ 12$&$ 12$\\
  \hline
$  99$&$(  2,  7,  9, 20, 31) $&$ 59$&$ 48$&$ 41$&$ 38$&$ 13$&$ 13$&
$ 100$&$(  2,  7, 10, 19, 31) $&$ 54$&$ 42$&$ 39$&$ 36$&$ 13$&$ 13$\\
  \hline
$ 101$&$(  1,  7,  9, 19, 28) $&$ 99$&$ 83$&$ 71$&$ 68$&$ 10$&$  9$&
$ 102$&$(  2,  3, 11, 24, 37) $&$116$&$111$&$ 83$&$ 80$&$ 12$&$ 12$\\
  \hline
$ 103$&$(  3,  5,  8, 24, 35) $&$ 64$&$ 47$&$ 43$&$ 40$&$ 10$&$  9$&
$ 104$&$(  3,  7, 10, 37, 54) $&$ 73$&$ 58$&$ 49$&$ 46$&$  9$&$ 10$\\
  \hline
$ 105$&$(  1, 12, 14, 40, 55) $&$119$&$113$&$ 94$&$ 91$&$  9$&$ 10$&
$ 106$&$(  4,  7,  9, 33, 46) $&$ 50$&$ 44$&$ 35$&$ 32$&$ 10$&$ 11$\\
  \hline
$ 107$&$(  5,  6,  7, 31, 44) $&$ 52$&$ 44$&$ 36$&$ 33$&$ 15$&$ 13$&
$ 108$&$(  3, 10, 18, 59, 87) $&$ 75$&$ 75$&$ 60$&$ 57$&$ 12$&$ 11$\\
  \hline
$ 109$&$(  3,  5,  9, 28, 42) $&$ 75$&$ 75$&$ 60$&$ 57$&$ 12$&$ 11$&
$ 110$&$(  2, 11, 14, 43, 59) $&$ 69$&$ 63$&$ 50$&$ 47$&$ 11$&$ 12$\\
  \hline
$ 111$&$(  1,  8, 18, 27, 51) $&$120$&$101$&$ 86$&$ 83$&$ 13$&$ 11$&
$ 112$&$(  1,  9, 20, 30, 57) $&$120$&$101$&$ 86$&$ 83$&$ 13$&$ 11$\\
  \hline
$ 113$&$(  1, 14, 18, 21, 27) $&$ 60$&$ 53$&$ 45$&$ 42$&$ 11$&$ 10$&
$ 114$&$(  1, 11, 14, 16, 21) $&$ 60$&$ 53$&$ 45$&$ 42$&$ 11$&$ 10$\\
  \hline
$ 115$&$(  2,  7, 24, 59, 85) $&$113$&$111$&$ 81$&$ 78$&$ 10$&$ 10$&
$ 116$&$(  4,  5, 39, 91,134) $&$115$&$110$&$ 81$&$ 78$&$  9$&$ 10$\\
  \hline
$ 117$&$(  3, 14, 15, 18, 25) $&$ 28$&$ 38$&$ 32$&$ 29$&$  9$&$ 11$&
$ 118$&$(  4,  4, 11, 17, 19) $&$ 34$&$ 34$&$ 24$&$ 21$&$  8$&$  8$\\
  \hline
  \end{tabular}
}
  \end{table}
\end{center}

\subsection{Three-Vector Chains}

Again as already mentioned, there are 259 three-vector chains
$m_1 {\vec k}^{(1,ext)} + m_2 {\vec k}^{(2,ext)} + m_3 {\vec
k}^{(3,ext)}$, where $m_1, m_2$ and $m_3$ are
integers. Also as mentioned earlier, the combinatorial problem of
checking each of the possible integer
combinations to identify those corresponding to reflexive polyhedra and
hence CY$_3$ spaces necessitated a short-cut solution. This was found in
the form of the `expansion' technique described above, whose
extension to the three-vector case we now describe.
To find the full list of projective vectors in  each of the 259
three-vector chains, one should distinguish two cases. Perhaps (a)
one can first find a non-trivial K3 intersection structure
from the list of
4~242 two-vector chains by making a simple expansion from a reflexive
polyhedron of dimension $D = 2$ to dimension $D + 1 = 3$, and then make a
second simple step, as described above, from $D + 1 = 3$ to dimension $D
+ 2 = 4$.
In other cases (b), one makes a double step directly from
$D = 2$ to dimension $D + 2 = 4$.

We do not dwell further on the two-step expansion (a), but illustrate
the double-step expansion (b) with a worked example. In this case, one
first identifies a common planar reflexive polyhedron of monomials
${\vec \mu}$,
and then chooses one of its vertices as a `double-expansion' point
$P$, leaving invariant the other vertices $V_i$. One then looks for
triples $P_1, P_2, P_3$ of points with
two components equal to zero and the property: 
\begin{eqnarray}
P \, = \, \frac{1}{3} (\, P_1\,+\,P_2\,+\,P_3)
\end{eqnarray}
that solve the following chain equations:
\begin{eqnarray}
{\vec k}_5\, . \, P_1\,&=&\,{\vec k}_5\, . \, P_2\,=\,{\vec k}_5\,
. \, P_3\,=\,d, \nonumber\\
{\vec k}_5\, . V_1\,&=&\,{\vec k}_5\, . \, V_2\,=\,d,
\end{eqnarray}
and then takes the union of these sets.

The 259 three-vector chains may be classified 
according to which of the 9 different planar reflexive polyhedra,
corresponding to $CP^2$ spaces, appears as a fibre.
One particular example is shown in Fig.~\ref{fig:vexp}(b): it is the
biggest $CY_3$ chain 
$m(1,0,0,2,3) + n(0,1,0,2,3)+l(0,0,1,2.3)=
(m,n,l,2m+2n+2l,3m+3n+3l)[6m+6n+6l]$, which has an 
elliptic Weierstrass fibre.
There are two fixed points in 
this chain: $V_1=(0,0,0,0,2)$ and $V_2=(0,0,0,3,0)$, and 
we expand around the point $P=(6,6,6,0,0)$ shown in
Fig.~\ref{fig:vexp}(b).
The eldest vector in this chain is $(1, 1, 1, 6, 9) [18]$, and
the youngest vector we find is $(91, 96, 102, 578, 867) [1734]$.
In total, we find 20~796 projective vectors
in this particular chain, but only 733 of them are new ones,
not found among the 4~242 two-vector chains~\footnote{We recall that a
similar overlap
was found among K3 spaces appearing in two- and three-vector
chains~\cite{AENV}.}.

The new values of the pairs of Hodge numbers found from another
thre-vector chain, namely $m (0, 0, 1, 1, 1) + n (0, 1, 0, 1, 2)
+ l (1, 0, 0, 1, 2)$, are shown in Fig.~\ref{fig:chains}(c).
The total number of distinct new projective vectors,
corresponding to new reflexive polyhedra in four dimensions and
hence CY$_3$ spaces, found
among
all the 259 three-vector chains is 6~189: their Hodge numbers
are plotted as circles in Fig.~\ref{fig:chains}(d). We recall that,
by construction, these new CY$_3$ spaces have elliptic {sections}.
Curiously, among all the three-vector chains,
we find no new CY$_3$ spaces with $N_g = 3$.

\subsection{Four-Vector Chains}

The above methods may be extended to four-vector chains
$m_1 {\vec k}^{(1,ext)} + m_2 {\vec k}^{(2,ext)}$ $+ m_3 {\vec
k}^{(3,ext)}
+ m_4 {\vec k}^{(4,ext)}$, where $m_1, m_2, m_3$ and $m_4$ are
integers. In this case, one may consider
the triple expansion directly 
from the reflexive line segment in dimension $D = 1$
if one expands about a single point $P$: 
\begin{eqnarray}
P\, = \, \frac{1}{4}(P_1\,+\,P_2\,+\,P_3\,+\,P_4),
\end{eqnarray}
one must solve the corresponding chain equations:
\begin{eqnarray}
&&{\vec k}_5 \, . \, P_1\,=\,{\vec k}_5 \, .\, P_2\,=\,{\vec k}_5 \, 
. \, P_3\,= \,{\vec k}_5 \, . \, P_4\,=\,d \nonumber\\
&&{\vec k}_5 \, . \, V_2\,=\,d. 
\end{eqnarray}
One example is the four-vector chain
$ (m,n,k,l m+n+k+l)[2m+2n+2k+2l]$, which one may expand
around the point $P=(2,2,2,2,0)$. The
eldest vector in this chain is $\vec{k}_5=(1,1,1,1,4)[8]$, and the
youngest vector is  $\vec{k}_5=(75,84,86,98,343)[686]$, whilst the
total number of projective vectors we find is 14~017.
The total number of new projective vectors found in all the
6 four-vector chains is 1~582. The new Hodge numbers of the corresponding
CY$_3$ spaces are shown as crosses in Fig.~\ref{fig:chains}(d).

\subsection{The Five-Vector Chain}

Finally, we note that the single five-vector chain
may be expanded about the point $P=(1,1,1,1,1)$:
\begin{eqnarray}
P \, = \, \frac{1}{5}(P_1+P_2+P_3+P_4+P_5).
\end{eqnarray}
it contains 7~269 projective vectors, of which 199 are new,
i.e., not contained among any of the
two-, three- or four-vector chains, none of which
have $N_g = 3$.

\section{Summary and Prospects}

We have presented in this paper some first results from our
systematic classification of CY$_3$ spaces,
building on our previous study of lower-dimensional
projective vectors, reflexive polyhedra and K3 spaces~\cite{AENV}.
Our approach gathers CY$_3$ spaces into chains composed from
extensions of lower-dimensional projective vectors, whose
fibration structure is explicitly apparent, as illustrated in the two
examples exhibited in Fig.~\ref{fig:vexp}. We have reported
the classification of 182~737 CY$_3$ spaces spanning 10~882
distinct values of the Hodge numbers $h_{11}, h_{12}$, as
seen in Figs.~\ref{fig:chains} and \ref{fig:histograms}.
Among these, we have reported the projective vectors,
Hodge numbers and other properties of 212 CY$_3$ with
three generations and K3 {sections}, shown in the Table.

It would be interesting to explore these three-generation models in
more detail, to see whether any of them are of potential
phenomenological interest. The information provided in this
paper is sufficient as a starting-point for such a programme of
work. The locations of the three-generation models in `chains' of
related CY$_3$ spaces may eventually
provide a tool for understanding transitions between different
string vacua, and the dynamical selection of the one we (presumably)
occupy. Such a study would require further information beyond that
provided in this paper, and we plan to provide the most important
data in a forthcoming more detailed paper~\cite{FrancoFTP}.

As was shown in~\cite{AENV} in the context of K3 spaces, our algebraic
approach opens the way to a systematic study of the gauge groups
obtainable at the singularities of CY$_3$ spaces. This is one of the
avenues for possible extensions of this work. Other major  
extensions include the application of our methods to CY$_4$
spaces, which are of interest in the context of $M$ theory. At
the same time, one can construct `higher-level' CY$_3$ spaces
defined as the intersections of polynomial loci. In the case of
K3 spaces, we found more spaces at the higher level than at the lowest
level~\cite{AENV}, and we expect a similar pattern for CY$_3$ spaces.

In conclusion, we are optimistic that the results presented here
and the prospects they reveal for the future of the `Calabi-Yau Genome
Project' will make possible
a much more informed and systematic assault on one of the central
problems of string phenomenology, namely the identity and
properties of the string vacuum.

\begin{center}

{\bf Acknowledgements}\\

{~~}\\

\end{center}

G.V. thanks E. Alvarez, R. Coquereaux, H. Dahmen, L. Fellin,
V. Petrov, A. Zichichi
and the CERN Theory Division for support. The work of D.V.N. was
supported in part by DOE grant no. DE-FG-0395ER40917.

\end{document}